\documentclass[12pt]{iopart}

\usepackage{bm}
\usepackage{amsfonts}
\usepackage{wasysym}
\usepackage{graphicx}
\usepackage{hyperref}
\usepackage{xcolor}
\usepackage{orcidlink}

\def\be{\begin{equation}}
\def\ee{\end{equation}}
\def\bea{\begin{eqnarray}}
\def\eea{\end{eqnarray}}
\def\vac#1{{\bf #1}}
\def\bsigma{{\boldsymbol\sigma}}
\def\half{{\textstyle \frac 12}}

\def\nnb{\nonumber}
\def\FigTexte#1#2#3#4#5{
\begin{figure}[ht]
\vskip#1mm
\begin{center}\includegraphics[scale=#2]{#3}\end{center}
\vskip#4cm
\caption{#5}
\end{figure}
}
\def\dd#1#2{_{#1#2}}

\begin{document}

\title[The advance of Mercury's perihelion]{The advance of Mercury's perihelion}

\author{Bertrand Berche$^1$  \orcidlink{0000-0002-4254-807X}, Ernesto Medina$^2$ \orcidlink{0000-0002-1566-0170}}

\address{$^1$ Laboratoire de Physique et Chimie Th\'eoriques, Universit\'e de Lorraine - CNRS, Nancy, France}

\address{$^2$ Departamento de F\'isica, Colegio de Ciencias e Ingenier\'ia, Universidad San Francisco de Quito, Diego de Robles y V\'ia Interoce\'anica, Quito, 170901, Ecuador}

\ead{bertrand.berche@univ-lorraine.fr}
\vspace{10pt}
\begin{indented}
\item[]\today
\end{indented}

\begin{abstract}
A very famous ``test''  of the General Theory of Relativity (GTR)  is the advance of Mercury's perihelion (and of other planets too). To be more precise, this is not a prediction of General Relativity, since the anomaly was known in the XIXth century, but no consistent explanation had been found yet at the time GTR was elaborated. Einstein came up with a solution to the problem in 1914.
 In the case of Mercury, the closest planet to the Sun, the effect is more pronounced than for other planets, and observed from Earth; there is an advance of the perihelion of Mercury of about 5550~arc seconds per century (as/cy). Among these, about $5000$ are due to the equinox precession (the precise value is {$5025.645$}~as/cy) and about $500$ ({$531.54$}) to the influence of the external planets. The remaining, about $50$~as/cy ({$42.56$}), are not understood within Newtonian mechanics.
Here, we revisit the problem in some detail for a presentation at the undergraduate level. 
\end{abstract}

%

\noindent{\it Keywords\/}: {Mercury, perihelion, Kepler problem, General Relativity}
%

\submitto{\EJP}
%
\maketitle
%
%

\section{Introduction}

The problem of the advance of Mercury's perihelion is a well-known example of a phenomenon that remained unexplained for a long time and which attracted the attention of many physicists in the 19th century before finding an interpretation within the framework of the theory of General Relativity. It is now a textbook case that played an important role in the acceptance of this theory and which is taught to illustrate the successes of General Relativity. The study of this problem is very rich because it allows us to illustrate the usefulness of perturbative calculations in classical Newtonian dynamics to understand the most important part of the effect observed as being due to the influence of external planets on the movement of Mercury, before devoting oneself to the relativistic approach of the gravitational effect due to the Sun which fully explains the residual advance, extremely tenuous, with a remarkable precision. However, the literature does not generally present what could be considered as the ultimate verification: how the influence of external planets, when treated in a relativistic framework, {\it does not bring} an additional correction, which could compete with the exceptional agreement between General Relativity and observations. The purpose of this article is to present this entire approach at an undergraduate level.

\section{The Kepler problem in Newton dynamics and the statement of the problem}\label{secKepler}

As a starting point, let us consider the Newtonian approach to planets' motion around the Sun.
This is an undergrad problem that can be found in all textbooks on classical mechanics~{\cite{LandauLifshitz,Goldstein}}. We look for bound states in the Newtonian gravitational potential of the Sun. The gravitational force is central, therefore
the angular momentum $\bsigma=\vac r\times m\vac v=mr^2\dot\theta\ \!\vac u_\varphi-mr^2\sin\theta\dot\varphi\ \!\vac u_\theta$ is conserved  (we use spherical coordinates, and the dot denotes a derivative w.r.t. time). The planet motion is thus confined to stay in a plane perpendicular to $\bsigma$ w.r.t. which the angle $\theta$ is measured, ($\theta$ is fixed to $\pi/2$) and $\bsigma=-mr^2\dot\varphi\vac u_\theta(\pi/2)=+mr^2\dot\varphi\vac u_z$ while $\vac v=\dot r\vac u_r+r\dot\varphi\vac u_\varphi$. The expression of the square of the velocity  $|\vac v|^2=\dot r^2+|\bsigma|^2/(m^2r^2)$ follows and is used in the second constant of motion, the total energy,
 \be
 E=\half m|\vac v|^2-\frac{GM_\odot m}{r}=\frac{|\bsigma|^2}{2{m}r^2}\Bigl( \frac 1{r^2}\Bigl( \frac{dr}{d\varphi} \Bigr)^2+1\Bigr)-\frac{GM_\odot m}{r}
 \ee
 where we also used $\dot r=\dot \varphi \ \! dr/d\varphi$. Here, $G=6.67430 \times 10^{-11} {\rm m}^3\ \! {\rm kg}^{-1}{\rm s}^{-2}$ is Newton's gravity constant and $M_\odot=1.9891 \times 10^{30}$kg the mass of the Sun, while $m=M_{\hbox{\mercury}}=3.285 \times 10^{23}$kg is the mass of Mercury (data for the planets of the solar system are listed in table~\ref{table-dataplanets}). We rewrite the energy equation in the form
 \be
 \frac 1{r^2} \Bigl(\frac{dr}{d\varphi} \Bigr)^2+1=\frac{2mr^2}{|\bsigma|^2}\Bigl( E+\frac{GM_\odot m}{r}\Bigr)
 \label{eq-Kepler_E}
 \ee
 and perform a change of variable $u=1/r$ to get
 \be
 \Bigl( \frac{du}{d\varphi}\Bigr)^2+u^2=\frac{2m}{|\bsigma|^2}( E+{GM_\odot m}u)\label{Kepler-1stIntegral-u}
 \ee
 which is then differentiated w.r.t. $\varphi$ to lead to a harmonic equation:
 \be
 \frac{d^2u}{d\varphi^2}+u=\frac{GM_\odot m^2}{|\bsigma|^2}.\label{Eq_KeplerLevel}
\ee
Adding a particular solution  of the complete equation, $u_0=\frac{GM_\odot m^2}{|\bsigma|^2}$, to a general solution $A\cos(\varphi-\alpha)$ of the homogeneous equation, and fixing the initial conditions to have $\alpha=0$, we obtain the solution 
\be
u(\varphi)=\frac{GM_\odot m^2}{|\bsigma|^2}(1+e\cos\varphi)\label{Kepler-solution-u}
\ee
where $e$ is called the eccentricity. The value of $e$ is fixed in terms of the constants  by the first equation of motion (\ref{Kepler-1stIntegral-u}) where the insertion of
$u(0)=u_0(1+e)$, $u'(0)=0$ provides the second order equation
$
u_0^2(1+e)^2-2u_0(1+e)-2mE/|\bsigma|^2=0.
$
We get closed ellipses when $E<0$ and the parameters of the trajectory are given by
\be
\frac 1r=\frac 1p(1+e\cos\varphi),\quad\frac 1p=u_0=\frac{GM_\odot m^2}{|\bsigma|^2},\quad e=\sqrt{1+\frac{2E|\bsigma|^2}{G^2M_\odot^2m^3}}.
\ee
In the case of Mercury, $e=0.205630$ and $p=a(1-e^2)=55.46\times 10^9$m with $a=57.91\times 10^9$m the semi-major axis.
Since $E~{\propto}~ m$ and $|\bsigma|~{\propto}~ m$, $p$ and $e$ do not depend on $m$; so the motion is completely fixed by the initial conditions, but doesn't depend on the planet mass $m$ (hence the name {\em universal} gravitation theory).

\begin{table}[ht]
\begin{center}
\begin{tabular}{ llll }
\br 
Planet \qquad\qquad& Mass (kg)\qquad & perihelion (km)\ \  & aphelion (km)   \cr 
\mr
\hbox{\rm Mercury}, \mercury & $3.302\times10^{23}$ & $46.0 \times 10^6$ & $69.8 \times 10^6$ \cr
 \hbox{\rm Venus}, \venus&  $4.868\times10^{24}$   & $107.5 \times 10^6$ & $108.9 \times 10^6$ \cr
 \hbox{\rm Earth}, \earth &  $5.974\times10^{24}$  & $147.1 \times 10^6$ & $152.1 \times 10^6$  \cr
 \hbox{\rm Mars}, \mars &  $6.418\times10^{23}$   & $206.6 \times 10^6$ & $249.2 \times 10^6$ \cr
 \hbox{\rm Jupiter}, \jupiter &  $1.899\times10^{27}$  & $740.7 \times 10^6$ & $816.1 \times 10^6$  \cr
 \hbox{\rm Saturn}, \saturn &  $5.685\times10^{26}$   & $1.349\times 10^9$ & $1.504 \times 10^9$ \cr
 \hbox{\rm Uranus}, \uranus &  $8.683\times10^{25}$  & $2.735 \times 10^9$ & $3.006 \times 10^9$  \cr
 \hbox{\rm Neptune}, \neptune &  $1.024\times10^{26}$  & $4.459 \times 10^9$ & $4.537 \times 10^9$  \cr
 \hbox{\rm Sun,} \astrosun &  $1.989\times10^{30}$ \cr
 \br
\end{tabular}
\caption{Masses and distances in the solar system.}
\label{table-dataplanets}
\end{center}
\end{table}

In conclusion, the previous approach assumes point-like masses describing Mercury and the Sun with Newtonian central forces operating; the elliptical orbit of Mercury is closed; thus, the perihelion always occurs at the same place in the orbital motion. In the literature this fixed orientation of the orbit is characterized by the conserved (fixed) Runge-Lenz vector ${\bf A}={\bf p}\times {\bf L}-mk\hat {\bf r}$ .

Now, since we know that the observations show that the motion indeed stays in a plane, but the ellipse is not closed in reality, its perihelion, the closest distance to the sun, rotates (actually, it precesses). In the case of Mercury, the closest planet to the Sun, this is more pronounced than for other planets. Observed from the Earth, there is a known advance of the perihelion of Mercury of  {$5599.745$}~arc seconds per century (as/cy) (the data are from Ref.~\cite{Clemence}), less than two degrees per century, which doesn't seem to be a very large amount, but astronomical observations were already very accurate long ago!  Of these, the equinox precession is responsible for most of the observed value,  {$5025.645$}~as/cy. This is the first and largest contribution to the perihelion advance, and it is because the Earth (from where we observe the phenomenon) is not perfectly spherical, and the combined effect of the Sun and the Moon exerts a torque on the Earth which produces a precession of the daily rotation axis of the Earth with a period of about $26000$ years, first observed by Hipparcos in 125 BC. The order of magnitude of these $\approx 26000$ years is thus $360\times 3600 / 260\simeq 5000$ as/cy. As said above, this is an order of magnitude, and the measurement is {$5025.645$}. This is not an anomaly of Mercury's motion but a relative effect due to the motion of the observer located on Earth.
 
The question we address now is about the remaining differences and the importance of the various causes we may invoke.

\section{The influence of external planets within the Newtonian theory of gravitation}\label{section2}

\begin{figure}[ht]
\begin{center}
\includegraphics[scale=0.80]{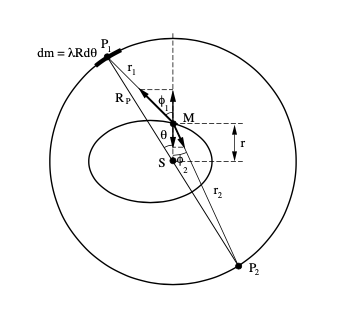}
\vskip0.0cm
\caption{Gravitational influence of an external planet P on the motion of Mercury (internal trajectory) in the solar system.}\label{fig-Planet-Mercury}
\end{center}
\end{figure}

The second effect is more difficult to analyze, even in Newtonian dynamics. We consider the effect on Mercury's motion of an external planet P of mass $M_{\rm P}$, the gravitational influence of which is assimilated to that of a ring of matter of radius $R_{\rm P}$, i.e., the mean radius of the orbit of P~\cite{PriceRush,Davies}, centered on the position of the Sun. {This is obviously an approximation, but we will see that the predictions are consistent in terms of orders of magnitude with more sophisticated calculations found in the literature. This model introduces a non-Newtonian ($1/r^2$) force effect, which breaks the conservation of the Runge-Lenz vector and makes it rotate, producing a precession of the perihelion.} This circular ring of matter carries a linear mass density of $\lambda=M_{\rm P}/(2\pi R_{\rm P})$. An important property here is that all the planets have their orbits almost on the same plane, so Mercury's position is in
the plane of planet P's orbit. If one considers two symmetric positions on the ``ring'', they both carry a mass element $dm=\lambda R_{\rm P}d\theta$
and produce on Mercury's position a gravitational field contribution which, after being projected on the direction from the Sun to Mercury, is written as 
\be
|d\vac G_{\rm P}|=G\lambda R_{\rm P}d\theta\Bigl(\frac {\cos\phi_1}{r_1^2}-\frac {\cos\phi_2}{r_2^2}\Bigr)
\simeq G\lambda d\theta\Bigl(\frac {1}{r_1}-\frac {1}{r_2}\Bigr)d\theta.
\ee
The angles and distances are defined in figure \ref{fig-Planet-Mercury}. The last equality comes from the approximation $R_{\hbox{\mercury}}\simeq p < R_{\rm P}$ (for example in the case of Venus, $R_{\rm P}$, of the order of the semi-major axis, is about $a_{\hbox{\venus}}=108.209\times 10^9$km), in which case $\phi_1\simeq\phi_2\simeq\theta$.
Then, with 
\bea
r_1=-r\cos\theta +\sqrt{r^2\cos^2\theta-(r^2-R_{\rm P}^2)},\\
r_2=+r\cos\theta +\sqrt{r^2\cos^2\theta-(r^2-R_{\rm P}^2)},
\eea
we get a (repulsive) central gravitational field after integration over $\theta$,
\be
\delta\vac G_{\rm P}=+\frac{G M_{\rm P}}{2R_{\rm P}}\frac{r}{R_{\rm P}^2-r^2}\vac u_r\simeq\frac{G M_{\rm P}}{2R_{\rm P}^3}\Bigl[ r+\frac{r^3}{R_{\rm P}^2}\Bigr]\vac u_r,
\ee
where the expansion is allowed because the radial distance of Mercury to the Sun $r< R_{\rm P}$, the radius of the orbits of the external planets (typically, for Mercury and Venus, the ratio $r/R_{\rm P}$ is about $0.5$ -- it could be worth taking into account the second-order expansion -- and the correction is smaller for the other planets).

It is convenient to write the gravitational potential due to the planet P:
\bea
\delta\phi_{\rm P}&=&-\int \delta\vac G_{\rm P}\cdot d\vac r\nnb\\
&=&\frac{G M_{\rm P}}{4R_{\rm P}}\ln\Bigl( 1-\frac{r^2}{R_{\rm P}^2} \Bigr)\simeq- \frac{G M_{\rm P}}{4R_{\rm P}^3}\Bigl[
r^2+\frac{r^4}{2R_{\rm P}^2}
\Bigr]
\eea
and the total gravitational potential energy (Sun plus the Planet P) as a correction to the simple effect of the Sun,
\bea V(r)&=&-\frac{GM_\odot m}{r}+\delta V_{\rm P}(r)
\nnb\\
&\simeq&
-\frac{GM_\odot m}{r}
- \frac{G M_{\rm P}m}{4R_{\rm P}^3}\Bigl[
r^2+\frac{r^4}{2R_{\rm P}^2}
\Bigr]
\quad\hbox{for $r < R_{\rm P}$}
\nnb\\
&=&-\frac{GM_\odot m}{r}\Bigl(
1+\frac{M_{\rm P}}{M_\odot}\frac{r^3}{4R_{\rm P}^3}
\Bigl[
1+\frac{r^2}{2R_{\rm P}^2}
\Bigr]
\Bigr). \label{eq-approx-Gpotential}
\eea

With the analytic form of the correction $\delta V_{\rm P}(r)$ to the purely Newtonian gravitational potential energy, it is possible to estimate the effect on the trajectory of Mercury using a perturbation analysis.

In Section \ref{secKepler}, we have established the equation of motion (\ref{Eq_KeplerLevel}) for the Kepler problem, i.e. the
motion of $m$ in the gravitational field exerted by the Sun. For a more general {\em central} potential corresponding now to Sun plus planet P,
\be
V(u)=-GM_\odot m u+\delta V_{\rm P}(u),
\ee
 the equation of motion becomes
 \be
 \frac{d^2u}{d\varphi^2}+u=
 -\frac m{|\bsigma|^2}\frac{dV(u)}{du}=
 \frac{GM_\odot m^2}{|\bsigma|^2}
  -\frac m{|\bsigma|^2}\frac{d\delta V_{\rm P}(u)}{du}
 .\label{Eq_Kepler+PLevel}
\ee
If we denote $u_{\rm K}$ the solution of Kepler problem and we seek for a perturbation solution, setting $u(\varphi)=u_{\rm K}(\varphi)+u_1(\varphi)$, the derivative of $\delta V_{\rm P}(u)$ w.r.t. $u$ at the r.h.s. is expanded in the vicinity of Kepler solution, and the equation of motion becomes
\bea
 \frac{d^2u_{\rm K}}{d\varphi^2}+u_{\rm K}+ \frac{d^2u_1}{d\varphi^2}+u_1=\frac 1p
 -\frac m{|\bsigma|^2} \left.\frac{d\delta V_{\rm P}(u)}{du}\right|_{u_{\rm K}}
-\frac m{|\bsigma|^2} \left.\frac{d^2\delta V_{\rm P}(u)}{du^2}\right|_{u_{\rm K}}u_1+\dots\nnb\\
\eea
Simplifying the equation at the Kepler problem level, we obtain an equation for the correction
\be  \frac{d^2u_1}{d\varphi^2}+\Bigl(1+\frac m{|\bsigma|^2} \left.\frac{d^2\delta V_{\rm P}(u)}{du^2}\right|_{u_{\rm K}}\Bigr)u_1
=
 -\frac m{|\bsigma|^2} \left.\frac{d\delta V_{\rm P}(u)}{du}\right|_{u_{\rm K}}
\ee
where we read that the angular frequency of the perturbed motion is
\be
\Omega^2=1+2\omega_1=1+\frac m{|\bsigma|^2} \left.\frac{d^2\delta V_{\rm P}(u)}{du^2}\right|_{u_{\rm K}},
\ee
hence the solution $u_1(\varphi)=A\cos(\Omega\varphi)+B\sin(\Omega\varphi)$.

The perihelion (the smallest value of $r(\varphi)=1/u(\varphi)$) corresponds to the largest value of $u(\varphi)$. It is obtained at $\varphi=0$ and equals to $u_{\rm max}=(1+e)/p+A$, ($A>0$). The same value is recovered slightly after a revolution at $\varphi=2\pi+\Delta\varphi$ (it will appear that $\Delta\varphi>0$) such that $u_{\rm K}(2\pi+\Delta\varphi)=(1+e)/p+O(\Delta\varphi^2)$ and $u_1(2\pi+\Delta\varphi)=A+B(2\pi(\Omega-1)+\Omega\Delta\varphi)+O(\Delta\varphi^2)$. To linear order in $\Delta\varphi$, the perihelion is recovered if the coefficient of $B$ vanishes. It follows that the advance (this will appear to be positive) of the perihelion per revolution due to the gravitational force exerted by the planet P is written as:
 \be
\Delta\varphi_{\rm P}=\frac{2\pi(1-\Omega)}{\Omega}
= -2\pi\omega_1
= -2\pi \frac 12\frac m{|\bsigma|^2} \left.\frac{d^2\delta V_{\rm P}(u)}{du^2}\right|_{u_{\rm K}}.\label{eq-shiftP}
 \ee
Using the expression of $\delta V_{\rm P}(u)$ we get to leading order
\be
\Delta\varphi_{\rm P}=2\pi \frac{3M_{\rm P}}{4M_\odot}\frac{p^3}{R_{\rm P}^3}
=\pi\frac{3}{2}
\frac{M_{\rm P} } {M_\odot}
\Bigl( 
\frac{a(1-e^2)} {R_{\rm P}}
\Bigr)^3.
\ee

In the case of the planet Venus (for $M_{\hbox{\venus}}/M_\odot={2.447}\times10^{-6}$ and $p/R_{\hbox{\venus}}={0.495}$, where $R_{\hbox{\venus}}$ is taken as the arithmetic mean of the distances to the aphelion and the perihelion), we get
\bea\fl
\Delta\varphi_{\hbox{\venus}}\simeq {1.399}\times10^{-6}
\Bigl(\frac{{\rm rad}}{{\rm rev}}\Bigr)\frac{360}{2\pi}\Bigl(\frac{\rm deg}{\rm rad}\Bigr)\frac{3600}{1}\Bigl(\frac{\rm sec}{\rm deg}\Bigr)\frac{1}{0.240}
\Bigl(\frac{\rm rev}{\rm year}\Bigr)\frac{100}{1}\Bigl(\frac{\rm year}{\rm century}\Bigr)
\nnb\\
\simeq {120}\hbox{ arcsec}/\hbox{century}.
\eea 
The next term in the potential (\ref{eq-approx-Gpotential}) adds another correction of ${0.571}\times10^{-7}$rad/rev, or $49.087$~as/cy, hence a total contribution of the influence of Venus at the first order perturbation expansion of $169$~as/cy. This is the strongest correction among the planets, the next one being due to Jupiter, which is more distant but far more massive.

Another method is used, e.g., in Ref.~\cite{LoYoungLee}, in terms of forces, but we have adapted it here in terms of potential energies.
More precise values have been determined numerically in an article in Am. J. Phys.\cite{Davies}
where we can find more accurate data. The largest contributions are from Venus (the closest planet) and Jupiter (the heaviest planet), and in the previous paper mentioned, Davies reports the numerical estimates for each planet, e.g., for Venus
$\Delta\varphi_{\hbox{\venus}}={273.30}$~as/cy and for Jupiter 
$\Delta\varphi_{\hbox{\jupiter}}={156.75}$~as/cy  while for the Earth,
$\Delta\varphi_{\hbox{\earth}}={91.49}$~as/cy and for Uranus
$\Delta\varphi_{\hbox{\uranus}}={0.14}$~as/cy.
These results are in better agreement with those of the specialized literature (see table~\ref{table-shiftplanets}) than ours, but the method that we employed is suitable for a presentation at the undergrad level.

\begin{table}[ht]
\begin{center}
\begin{tabular}{ lll }
\br 
Origin &  $\Delta\varphi$ (as/cy)  & Ref.~\cite{NobiliWill}\cr 
\mr
 \hbox{\rm Venus}, \venus& $277.856\pm 0.68$ \cr
 \hbox{\rm Earth}, \earth & $90.038\pm 0.08$ \cr
 \hbox{\rm Mars}, \mars & $2.536\pm 0.00$ \cr
 \hbox{\rm Jupiter}, \jupiter & $153.584\pm 0.00$ \cr
 \hbox{\rm Saturn}, \saturn & $7.302\pm 0.01$ \cr
 \hbox{\rm Uranus}, \uranus & $0.141\pm 0.00$ \cr
 \hbox{\rm Neptune}, \neptune & $0.042\pm 0.00$ \cr
 \hbox{\rm Sun (\astrosun) asphericity} & $0.010\pm 0.02$ \cr
 \hbox{\rm general precession of the equinoxes} & $5025.645\pm 0.50$ \cr
\mr
 \hbox{\rm Sum} & $5557.18\pm 0.85$ \cr
 \hbox{\rm observed advance} & $5599.74\pm 0.41$ \cr
\mr
 \hbox{\rm remaining difference} & $42.56\pm 0.94$ \cr
  \hbox{\rm GTR effect} & $43.03\pm 0.03$ &$42.98$  \cr

 \br
\end{tabular}
\caption{Various contributions to the advance of the perihelion of Mercury, from G.M. Clemence~\cite{Clemence}. The GTR main contribution (final value, according to Eq.~(\ref{eq_GTRMain})) was later corrected by Nobili and Will~\cite{NobiliWill}.}
\label{table-shiftplanets}
\end{center}
\end{table}

When one sums up all the contributions, as well as a tiny effect due to the non-exact Newtonian form of the Sun's gravitational potential due to its non-perfect sphericity, there remains a very small difference with the observations. That difference is not explained by the classical theory of gravitation, as one can read in the results~\cite{Clemence} given 
in table \ref{table-shiftplanets}.

\section{Looking for possible explanations of the remaining $43$ arc seconds per century}

This tiny number, about $43$ arc seconds per century (compared to $~5599$ as/cy observed), has to find a further explanation.
A slight modification to Newton's law of gravitation has been proposed as well as the idea of a still unknown celestial object (which was even given a name, Vulcan), the influence of which would add to the other planets to produce the desired  $43$ arc seconds per century,
but nothing was discovered as a possible candidate. 
 
 A similar hypothesis had been put forward earlier to explain the anomalies in the motion of the planet Uranus. It was all the merit and glory of Le Verrier to specify by calculation the mass and position of the new planet, named Neptune, which was observed then by Galle in Berlin\footnote{The planet was within $1^\circ$ of where Le Verrier had predicted (and $10^\circ$ of where Adams had sooner predicted).}. In the XIXth century, carrying out perturbative calculations was not an easy task. A page of Le Verrier's calculations is given in 
 figure~\ref{Fig_Neptune}.
 
   \FigTexte{1}{0.70}{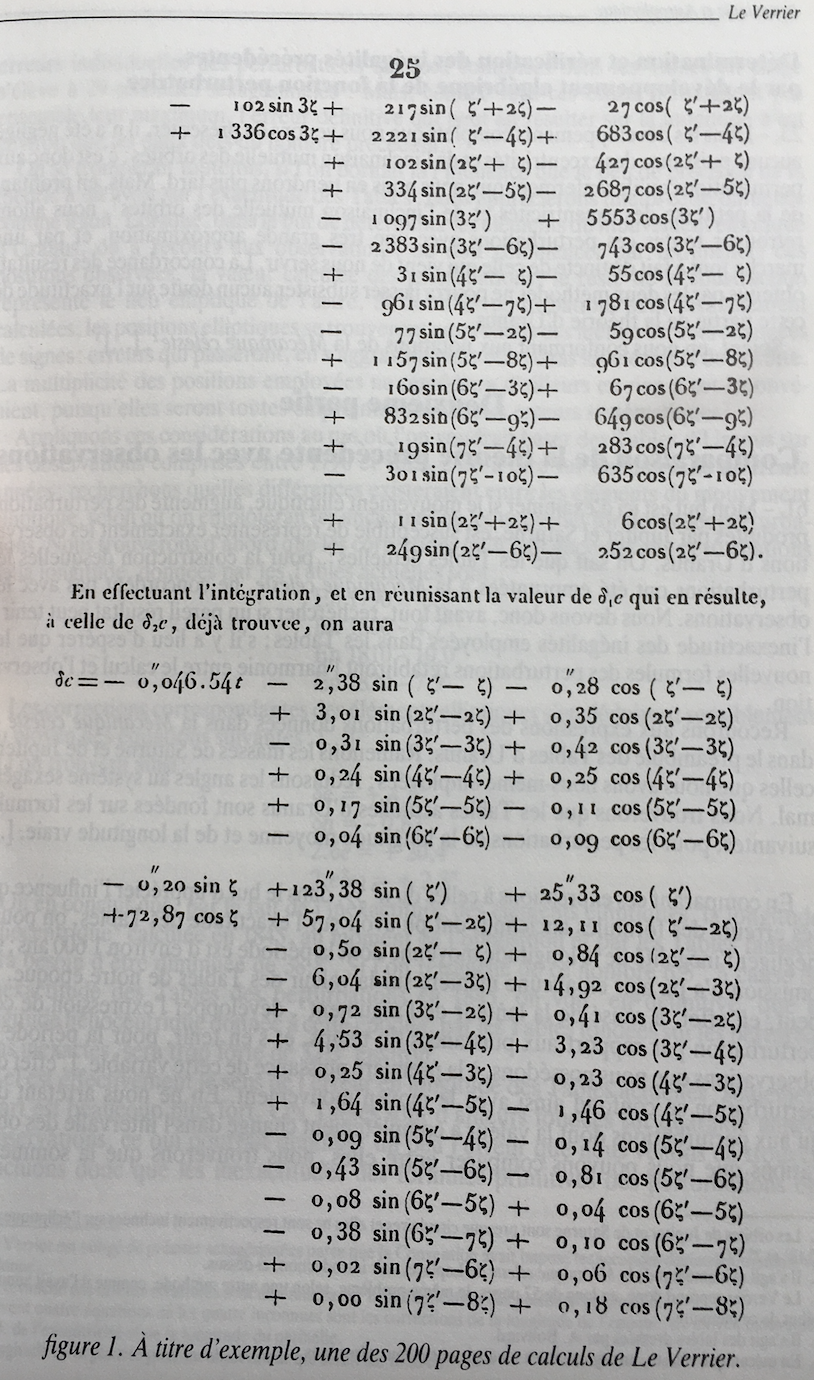}{-0.2}
{One of the 200 pages of calculation of Le Verrier. From J.-P. Verdet, Astronomie et astrophysique, Larousse, Paris 1993, p.731. \label{Fig_Neptune}}

 It is also instructive to read Le Verrier himself about the motion of Uranus~\cite{Verdet}:
 \begin{quotation}
 {\small\sf
A few years ago, we had barely begun to suspect that the movement of Uranus was modified by some unknown cause when all possible hypotheses were already hazarded on the nature of this cause. It is true that everyone simply followed the inclination of their imagination without providing any consideration to support their assertion. We thought of the resistance of the ether, we spoke of a large satellite which accompanied Uranus, or of a still unknown planet whose disturbing force should be taken into consideration, we even went so far as to suppose that at this enormous distance from the Sun, the law of gravitation could lose something of its rigour.
}
\end{quotation}

A modification of the law of gravitation at large distances is still a very current debate today.
 In the context of Mercury's anomalies, this hypothesis was  proposed as
 a correction to Newton's gravitation, considered by Hall \cite{Tonnelat,Roseveare}, who had shown that any $n>2$ in a gravitational force of the form $GMm/r^n$ would result in an advance of the perihelion, and that $n=2.000\ \! 000\ \! 16$
was enough to explain the mysterious $43$ as/cy of Mercury. But then, the same $n$ was spoiling the results concerning the other planets in the solar system, which was not acceptable.
 
 The assumptions mentioned by Le Verrier to explain Uranus anomalies are still among the most popular possible causes introduced in cosmology to explain the deviations observed in the evolution of the scale parameter of the metric of the Universe when only observable sources of gravity are considered. Dark matter is indeed similar to the introduction of a supplementary planet, unknown at the time, and dark energy can be considered the analogue of a modification of the law of gravitation.

Eventually, the route of a modification of gravity will appear successful for Mercury's anomaly. Indeed, there is no escape and a relativistic approach has to be used to try to solve the ``tiny $43$ arc seconds per century''. 
  
 \section{The resort to Special Relativity}
 
First the contribution of Special Relativity should be considered, and
there has been a controversy on the role of the purely special relativistic contribution, as one can see in this ``ironic'' quotation~\cite{McDonald}\footnote{We keep the reference numbering of the original quotation.}:

 \begin{quotation}
{\small
\sf The question arises as to what is the prediction from ``Special Relativity''. The literature on
this is rather erratic.

Early work (1906-1911) by Poincar\'e~\cite{21,22}, 
 Lorentz \cite{23}, de Sitter \cite{24} and others (...) inferred that the result from Special Relativity for the precession of the perihelion of Mercury is only $1/6$ that of the observed value. An effort by Nordstr\"om in 1912 \cite{25} predicted precession $-1/6$ of the observed value.

In 1917, Lodge \cite{26} claimed to be inspired by Special Relativity to consider velocity- dependent corrections to the precession of the perihelion, but actually reverted to Newton’s analysis of precession in case of a force law $1/r^n$ for $n$ different than $2$ \cite{27}, as extended by \cite{29, 30}. A debate followed between Eddington and Lodge \cite{31,32,33,34,35}.

In 1929, Kennedy \cite{36} gave two analyses of Newtonian precession of the perihelion, with corrections for retardation and for Special Relativity, claiming negligible effects in both cases. It was stated by Goldstein 
 \cite{37} (1950) that the result from
Special Relativity is $1/6$ that of General Relativity (...).

In 1984, Phipps \cite{39} claimed that the result of Special Relativity is the same as that of
General Relativity.

In 1986, Peters \cite{40} noted that Phipps made a computational error, and claimed the
correct result of Phipps’ model is $1/2$ that of General Relativity (\dots).

In 1987, Biswas \cite{41} claimed that the result of (his interpretation of) Special Relativity is the same as that of General Relativity.

In 1988, Frisch \cite{42} discussed ``post-Newtonian'' approximations, claiming that use of ``relativistic momentum'' but Newtonian gravity gives the result of Goldstein \cite{37}, $1/6$ of the observed precession of the perihelion of Mercury, while including the gravitation due to gravitational field energy doubles the result, to $1/3$ of the observed precession of the perihelion of Mercury.

In 1989, Peters \cite{43} argued that Biswas’ calculation was in error.

In 2006, Jefimenko proposed a theory of ``cogravitation'', and claimed it predicted $1/3$ of observed precession of the perihelion of Mercury (\dots).

In 2015, Wayne \cite{46} claimed that Special Relativity can explain the precession of the
perihelion of Mercury.

In 2016, Lemmon and Mondragon \cite{47} argued that Special Relativity predicts $1/3$ of the
rate of the precession of the perihelion according to General Relativity.

In 2020, Corda \cite{48} claimed that Newtonian gravity completely explains the precession of the perihelion of Mercury (without consideration of relativity), but not that of other planets. Then, he argued that General Relativity also explains the precession, but only if one includes
the effect of ``rotational time dilation''.

In 2022, D'Abramo \cite{49} claimed that Corda \cite{48} was wrong. 

What is going on here?
}
\end{quotation}

 To understand thecontroversy, let us first look at the purely kinematic contribution of Special Relativity (neglecting the influence of external planets).  The Lagrangian of a particle in a potential $V(r)=-GM_\odot m/r$ is
 \be
 L=-\frac {mc^2}\gamma+\frac{GM_\odot m}r
 \ee
 where $\gamma^{-1}=\sqrt{1-|\vac v |^2/c^2}$ and $|\vac v|^2=\dot r^2+r^2\dot\theta^2+r^2\sin^2\theta\dot\varphi^2$. Lagrange equations are therefore
 \bea
&& \gamma r\dot\theta^2+\gamma r\sin^2\theta\dot\varphi^2-\frac{GM_\odot m}{r^2}=\gamma\ddot r+\dot\gamma\dot r,\nnb\\
&&\gamma r^2\sin 2\theta=\frac d{dt}(\gamma r^2\dot\theta),\nnb\\
&&\frac d{dt}(\gamma r^2\sin^2\theta\dot\varphi)=0.
 \eea
Using the Lagrange equations is an option, but we could also use the first integrals, conservation of the angular momentum, and energy, as we did in Newtonian mechanics. The same works here, which leads to a more direct derivation.
 
The angular momentum is written as $\bsigma = \vac r\times m\gamma \vac v=\vac r\times\vac p$, so that $\frac {d\bsigma}{dt}=\vac v\times \vac p+\vac r\times\frac {d\vac p}{dt}$. The first term obviously vanishes, and the second vanishes for central potentials for which 
 $\frac{d\vac p}{dt}=\vac F=F(r)\vac r/r$.
 Now, since $\bsigma$ is conserved, we deduce again that the motion stays within a plane, and we choose $\theta=\pi/2$ ($\dot\theta=0$), measuring the $\theta$ angle w.r.t. the direction of the angular momentum,
 the kinematic factor $\gamma$ takes the form
 \be
 \gamma^2=\Bigl[ 
 1-\frac{1}{c^2}\left( \frac{dr}{d\varphi}\Bigr) ^2\dot\varphi^2-\frac {r^2}{c^2}\dot\varphi^2
 \right]^{-1}.\label{eq-gamma2}
 \ee
 A first constant of motion follows from the fact that $L$ doesn't depend on $\varphi$, hence
 \be
 |\bsigma|=p_\varphi=\frac{\partial L}{\partial\dot\varphi}=m\gamma r^2\dot\varphi.
 \ee
 The second constant of motion is obviously the energy, 
 \be\fl 
 E=p_r\dot r+p_\varphi\dot\varphi - L=m\gamma\dot r^2+m\gamma r^2\dot\varphi^2+mc^2\gamma^{-1}-\frac{GM_\odot m}{r}=m c^2\gamma-\frac{GM_\odot m}{r}.
 \ee
Using the definition of the angular momentum, we eliminate $\dot\varphi$ in (\ref{eq-gamma2}), 
and factorize out $\gamma^2$, leading to
 \be
 \gamma^2=1+\frac{|\bsigma|^2}{m^2c^2r^4} 
 \Bigl( \frac{dr}{d\varphi}\Bigr) ^2
 +\frac{|\bsigma|^2}{m^2c^2r^2}=\Bigl(
 \frac{E}{mc^2}+\frac{GM_\odot}{rc^2}
 \Bigr)^2 .
 \ee
 Like in the Newtonian case, the change of variable $u=\frac 1r$ simplifies the equation into
 \be
 1+\frac{|\bsigma|^2}{m^2c^2}\Bigl[
 \Bigl(
 \frac{du}{d\varphi}
 \Bigr)^2+u^2
 \Bigr]=\Bigl(
 \frac{E}{mc^2}+\frac{GM_\odot  u}{ c^2}
 \Bigr)^2 .\label{eq_SR_PotEff}
 \ee

  A formula analogous to (\ref{Eq_Kepler+PLevel}) is obtained if we write
 the equation of motion which follows from the derivative of (\ref{eq_SR_PotEff}) w.r.t. $\varphi$
 \be
 \frac{d^2u}{d\varphi^2}+\left(
1-\frac{G^2M_\odot^2 m^2}{|\bsigma|^2 c^2}
 \right) u
=\frac{GM_\odot m^2}{|\bsigma|^2}\left(
\frac{E}{mc^2}
\right). \label{Eq_SRLevel}
 \ee
 This equation
qualitatively differs from the classical case by the value of the angular velocity, which is now
\be \Omega^2=1-\frac{G^2M_\odot^2 m^2}{|\bsigma|^2 c^2} \simeq 1+2\omega_1\ee
that differs from unity, inducing a shift of the perihelion. 
 We can estimate this shift as we did for the influence of external planets.
 \be
 \Delta\varphi_{\rm SR}\simeq \frac{2\pi (1-\Omega)}{\Omega}=
 -2\pi\omega_1=
 2\pi\frac 12 \frac{G^2M_\odot^2 m^2}{|\bsigma|^2 c^2} 
 =\frac{\pi GM_\odot}{a(1-e^2)c^2}.
 \ee 
  The numerical value is estimated by inserting the classical parameters of the ellipse, $|\bsigma|^2=GM_\odot m^2 p$ and $p=a(1-e^2)$. 
  We get
  \be
   \Delta\varphi_{\rm SR}=8.649\times 10^{-8}\hbox{rad}/\hbox{rev}=7.433\ \!\hbox{as/cy}.\label{eq-SRcalc}
  \ee
To link with the quotation of McDonald's,  it appears to be
 6 times smaller than the observed one of 43. So we have to conclude that Special Relativity is not enough to explain the whole effect observed. 
However, we can  notice that in the controversy reported by McDonald~\cite{McDonald}, the pioneers Poincar\'e, Lorentz or de Sitter were right!

\section{How General Relativity solves the problem}

Now, let us follow the same lines of reasoning in full General Relativity. General Relativity encodes gravitational energy in the metric, 
\be
ds^2=(1-(2GM_\odot/(rc^2))c^2dt^2
-(1-(2GM_\odot/(rc^2))^{-1}dr^2-r^2d\Omega^2,\label{Eq-SchwMetric}
\ee
and the zero-mass limit recovers the case of Special Relativity, therefore there will be no need to add the result (\ref{eq-SRcalc}) to the present calculation.

The Lagrangian\footnote{The action of a free particle in Special Relativity $S=-mc\int ds$.} $L=-mc \frac{ds}{dt}$
now reads as
\be
L=-mc^2\frac{d\tau}{dt}=-mc^2\Bigl[
g\dd00(r)-\frac{1}{g\dd 00(r) c^2}\Bigl(\frac{dr}{dt}\Bigr)^2
-\frac{r^2}{c^2}\Bigl(\frac{d\varphi}{dt}\Bigr)^2
\Bigr]^{1/2}
\ee
in terms of the proper time $\tau$, and the potential term is hidden in the metric tensor component. The argument for planar motion still works. Hence we have already simplified the problem considering fixed $\theta=\pi/2$, and we have assumed Schwarzchild metric (\ref{Eq-SchwMetric})
$g\dd00(r)=1-\frac{2GM_\odot}{rc^2}=-1/ g\dd 11(r)$. Like in the special relativistic case, we have two constants of motion.
The angular momentum is the first
\be
|\bsigma|=\frac{\partial L}{\partial\dot\varphi}=mr^2\Bigl(
\frac{d\varphi}{d\tau}
\Bigr).\label{eq_consJGTR1}
\ee
Here, $\dot\varphi=d\varphi / dt$ shouldn't be confused with $d\varphi / d\tau$.
The momentum associated to the radial coordinate is equal to
\be
\frac{\partial L}{\partial\dot r}=\frac{m}{g\dd 00(r)}\Bigl(
\frac{d r}{d\tau}
\Bigr),
\ee
with the same notation $\dot r=dr / dt$. The energy follows
\be
E = \frac {d\tau}{dt}\Bigl[
\frac{m}{g\dd00(r)}\Bigl(\frac {dr}{d\tau}\Bigr)^2+\frac{|\bsigma|^2}{mr^2}+mc^2
\Bigr]
.
\label{eq_consEGTR1}
\ee
The square of the interval provides an alternative identity,
\bea
c^2 
&=& 
g\dd 00(r)\Bigl(\frac{cdt}{d\tau}\Bigr)^2
-\frac{1}{g\dd 00(r)}\Bigl(\frac{dr}{d\tau}\Bigr)^2
-r^2\Bigl(\frac{d\varphi}{d\tau}\Bigr)^2 ,
\label{eq_mc2GTR1}
\eea
which leads to the relation
\be
\frac{1}{g\dd 00(r)}\Bigl(\frac{dr}{d\tau}\Bigr)^2=
g\dd 00(r)c^2\Bigl(\frac{dt}{d\tau}\Bigr)^2 - c^2 -\frac{|\bsigma|^2}{m^2r^2}.
\ee
This latter expression is now
inserted in (\ref{eq_consEGTR1})
and one obtains the simple form
\be
E=g\dd 00(r)mc^2\Bigl(\frac{dt}{d\tau}\Bigr). \label{eq_consEGTR2}
\ee
Now, equations (\ref{eq_consJGTR1}) and (\ref{eq_consEGTR2}) inserted in (\ref{eq_mc2GTR1}) lead to
\bea
g\dd00(r)m^2c^2&=&\frac{E^2}{c^2}-\frac{|\bsigma|^2}{r^4}\Bigl(\frac{dr}{d\varphi}\Bigr)^2-g\dd00(r)\frac{|\bsigma|^2}{r^2}\label{eq-variable-r-GTR}\\
g\dd00(u)m^2c^2&=&\frac {E^2}{c^2}-|\bsigma|^2\Bigl(
\frac{du}{d\varphi}
\Bigr)^2
-g\dd 00(u) |\bsigma|^2 u^2\label{Eq-GTRKepler}
%
\eea
where the second line follows from the previous one by the usual change of variable $u=\frac 1r$. We next take another derivative w.r.t. $\varphi$ to get the
relativistic equation of motion
\bea
\frac{d^2 u}{d\varphi^2}+g\dd00 (u)u&=&-\frac12\frac{dg\dd00(u)}{du}\Bigl(\frac{m^2c^2}{|\bsigma|^2}+u^2
\Bigr),\label{Eq_GTRMetricLevel}\\
\frac{d^2 u}{d\varphi^2}+u&=&\frac{GM_\odot m^2}{|\bsigma|^2}+\frac{3GM_\odot}{c^2} u^2. \label{Eq_GTRLevel}
\eea
It is instructive to compare with the equation of motion in the Kepler approximation (\ref{Eq_KeplerLevel}) or in the special relativistic case (\ref{Eq_SRLevel}), both of them being linear. We rewrite these equations (with labels K for Kepler and SR for Special Relativity) here for the purpose of comparison,
\bea
\frac{d^2 u_{\rm K}}{d\varphi^2}+u_{\rm K}&=&\frac{GM_\odot m^2}{|\bsigma|^2},\\
\frac{d^2 u_{\rm SR}}{d\varphi^2}+u_{\rm SR}&=&\frac{GM_\odot m^2}{|\bsigma|^2}+\frac{G^2M_\odot^2 m^2}{|\bsigma|^2 c^2}u_{\rm SR}
\eea
and clearly the main difference is that in General Relativity we get a non linear equation. This latter equation in (\ref{Eq_GTRLevel}) is solved perturbatively around the classical case, allowing the periodic solution and harmonics at multiple frequencies, together with a possible shift of the fundamental frequency of the Kepler solution. We thus allow
\bea
&&\phi=\Omega\varphi=(1+\omega_1+\dots)\varphi,\\
&&u(\varphi)=u_{\rm K}(\phi)+u_1(\phi)+\dots,
\eea
where $\omega_1$ and $u_1$ are small perturbations,
with $u_{\rm K}(\phi)\sim\cos\phi$ and $u_1(\phi)\sim\cos2\phi$, \dots Inserting these expansions in  (\ref{Eq_GTRLevel}) leads to 
\bea
&&\Omega^2\frac{d^2 u}{d\phi^2}+u=\frac 1p+\frac{3GM_\odot}{c^2}u^2,\\
&0\hbox{th order\quad }&\frac{d^2 u_{\rm K}}{d\varphi^2}+u_{\rm K}=\frac 1p\\
&1\hbox{st order }&\frac{d^2 u_1}{d\varphi^2}+u_1=\hbox{const}+\frac{2e}{p}\Bigl(
\frac{3GM_\odot}{pc^2}+\omega_1
\Bigr)\cos\phi\nnb\\
&&\qquad\qquad\qquad\qquad +\frac{3GM_\odot e^2}{2p^2c^2}\cos 2\phi.
\eea
The 0th order solution is indeed Kepler solution and, demanding that $u_1$ has no dependence at the same frequency leads to the first order correction of the angular velocity,
\be
\omega_1=-\frac{3GM_\odot}{pc^2}.
\ee
It follows an advance of the perihelion
  \be
  \Delta\varphi_0=-2\pi\omega_1=\frac{\pi 6G^2M_\odot^2m^2}{c^2|\bsigma|^2}=\frac{6\pi GM_\odot}{a(1-e^2)c^2},
  \ee
 being 6 times larger than in the special relativistic treatment. 
 
 The numerical value is 
  \be
   \Delta\varphi_0=5.05\times 10^{-7}\hbox{rad}/\hbox{rev}
   =43.2\ \!\hbox{as/cy}\label{eq_GTRMain}
  \ee
  for Mercury.
This numerical result agrees remarkably with the observations collected, for the case of Mercury, in table~\ref{table-shiftplanets}. 
The results for other bodies in the solar system are given in table~\ref{table-shiftallplanets}

\begin{table}[ht]
\begin{center}
\begin{tabular}{ lll }
\br 
Celestial body &  $\Delta\varphi_0$ measured (in as/cy) & predicted in GTR  (as/cy) \cr 
\mr
 {\rm Mercury}& $43.11 \pm 0.45$ & $43.03$~\cite{Clemence} (or 42.98~\cite{NobiliWill})\cr
 {\rm Venus}& $8.4\pm 4.8$ & $8.6$\cr
 {\rm Earth}& $5.0\pm 1.2$ & $3.8$\cr
 {\rm Icarus}& $9.8\pm 0.8$ & $10.3$\cr
 \br
\end{tabular}
\caption{Advance of the perihelion in the solar system. From Ref.~\cite{French}. }
\label{table-shiftallplanets}
\end{center}
\end{table}

Resolving a disagreement between theory and observations that had been misunderstood for years was a major breakthrough that consolidated the General Theory of Relativity of Einstein. Although numerically minor, solving the discrepancy was essential on fundamental grounds and this opened the era of high precision astronomy and astrophysics.

\section{But what about the influence of external planets in GTR?}

For consistency, 
we must now ensure that taking into account the effect of the external planets {\em at the level of General Relativity} does not ruin the formidable agreement of the previous calculation. For that, let us remind that at the level of Newtonian gravitation theory,  the equation of motion for central potentials $V(r)$ reads as
\be
\frac{d^2u}{d\varphi^2}+u(\varphi)=-\frac m{|\bsigma|^2}\frac{dV}{du},
\ee
while in GTR the corresponding equation is (\ref{Eq_GTRMetricLevel})
\be
\frac{d^2 u}{d\varphi^2}+g\dd00 (u)u=-\frac12\frac{dg\dd00(u)}{du}\Bigl(\frac{m^2c^2}{|\bsigma|^2}+u^2
\Bigr)
\ee
with 
\be
g\dd00(u)=1+\frac{2\phi(u)}{c^2}.
\ee
Using the potential energy found in Section~\ref{section2}, limited to the leading order for the planet contribution,
\be
V(u)=m\phi(u)=-GM_\odot mu-\frac{GM_{\rm P}m}{4R_{\rm P}^3u^2},
\ee
 we get the following equation of motion in the classical case
\bea
\frac{d^2u}{d\varphi^2}+u(\varphi)&=&\underbrace{\frac {GM_\odot m^2}{|\bsigma|^2}}_{{\hbox{\footnotesize Sun influence \vrule height 0mm depth 1.2mm width 0pt}}\atop {\hbox{\footnotesize Newtonian level}}}-\underbrace{\frac{GM_{\rm P}m^2}{2R_{\rm P}^3|\bsigma|^2}\frac 1{u^3}}_{{\hbox{\footnotesize External planet influence \vrule height 0mm depth 1.2mm width 0pt}}\atop {\hbox{\footnotesize at the Newtonian level}}}
=\frac 1p-\frac{M_{\rm P}}{2M_\odot}\frac 1{pR_{\rm P}^3}\frac 1{u^3} .\nnb\\\label{Eq63}
\eea
This is the equation that we analyzed earlier. 

Within General Relativity, we have seen that the equation of motion takes the form of Eq.~(\ref{Eq_GTRMetricLevel}) with now
\bea
g\dd 00(u)=
1-\frac{2GM_\odot}{c^2}u-\frac{GM_{\rm P}}{2R_{\rm P}^3c^2}\frac{1}{u^2}.
\eea
The equation of motion follows in the form
\bea
\frac{d^2u}{d\varphi^2}+u(\varphi)&=&
\underbrace{\frac {GM_\odot m^2}{|\bsigma|^2}}_{{\hbox{\footnotesize Sun influence \vrule height 0mm depth 1.2mm width 0pt}}\atop {\hbox{\footnotesize Newtonian level}}}+
\underbrace{\frac {3GM_\odot}{c^2}u^2}_{{\hbox{\footnotesize Sun correction \vrule height 0mm depth 1.2mm width 0pt}}\atop {\hbox{\footnotesize GTR level}}}
\nnb\\ &&
-
\underbrace{
\frac{M_{\rm P}}{2M_\odot}\frac 1{pR_{\rm P}^3}\frac 1{u^3}
}_{{\hbox{\footnotesize External planet influence \vrule height 0mm depth 1.2mm width 0pt}}\atop {\hbox{\footnotesize at the Newtonian level}}}
-
\underbrace{
\frac{GM_{\rm P}}{4R_{\rm P}^5c^2}\frac 1{u^3}
}_{{\hbox{\footnotesize External planet correction \vrule height 0mm depth 1.2mm width 0pt}}\atop {\hbox{\footnotesize at the GTR level}}}\nnb\\ \label{Eq67}
\eea
Comparing Eq.~(\ref{Eq67}) with Eq.~(\ref{Eq63}) we see that only the last term differs in the planet contribution. When we compare the order of magnitude of the two contributions due to the planet, the GTR vs the classical contributions, the ratio is about $10^8$ times smaller.
This proves that there is no need for further analysis of this latter effect which leads, at the experimental accuracy, to the same results as in Newtonian gravity.
The 43 as/cy are indeed due to the sole correction of the influence of the Sun in General Relativity.

\section{A problem recently revisited}

C.M. Will is a specialist of the experimental verifications of the General Theory of Relativity. He recently revisited the problem of the advance of Mercury's perihelion~\cite{Will2018} and found a new contribution arising from the interaction between Mercury’s motion and the gravitomagnetic field of the moving planets. The numerical contribution is very small, with a few parts in a million of the GTR main contribution to precession, but considers that this should be detectable experimentally.
It has to be noticed that the GTR correction to the planets' contribution that we have estimated is still a factor 10 smaller, but it might be accessible in the future.

 \end{document}